\documentclass[twocolumn,aps,prl,floatfix,showpacs,amsmath,amssymb]{revtex4}
\usepackage{graphicx}
\usepackage{dcolumn}
\usepackage{bm}
\usepackage{threeparttable}
\usepackage{multirow,array}
\usepackage{makecell}

\begin{document}

\title{Critical Behavior of a Strongly Disordered 2D Electron System: \\ The Cases of Long-Range and Screened Coulomb Interactions}

\author{Ping V. Lin}
\email{lin@magnet.fsu.edu}\affiliation{National High Magnetic Field Laboratory, Florida State University, Tallahassee, Florida 32310, USA}

\author{Dragana Popovi\'c}
\email{dragana@magnet.fsu.edu}\affiliation{National High Magnetic Field Laboratory, Florida State University, Tallahassee, Florida 32310, USA}

\date{\today}

\begin{abstract}
 A study of the temperature ($T$) and density ($n_s$) dependence of conductivity $\sigma(n_s,T)$ of a highly disordered, two-dimensional (2D) electron system in Si demonstrates scaling behavior consistent with the existence of a metal-insulator transition (MIT).  The same critical exponents are found when the Coulomb interaction is screened by the metallic gate and when it is unscreened or long range.  The results strongly suggest the existence of a disorder-dominated 2D MIT, which is not directly affected by the range of the Coulomb interactions. 
\end{abstract}


\maketitle

The metal-insulator transition (MIT) in 2D systems remains one of the most fundamental open problems in condensed matter physics \cite{2DMIT-review_2001, 2DMIT-review_2004, 2DMIT-review_2010}.  There is considerable experimental evidence that suggests that electron-electron interactions are responsible for a variety of phenomena observed in the metallic regime of low-disorder 2D systems near the apparent MIT, including a large increase of conductivity $\sigma$ with decreasing temperature $T$ ($d\sigma/dT<0$) \cite{Radonjic}.  Many-body effects have been most pronounced in a 2D electron system (2DES) in Si metal-oxide-semiconductor field-effect transistors (MOSFETs).  Since the most striking
experimental features are not sensitive to weak disorder (see, \textit{e.g.}, the thermopower study in Ref.~\cite{Kravchenko-thermo}),  they have been interpreted as evidence that the MIT in such low-disorder systems is driven by electron-electron interactions and that disorder has only a minor effect.  In highly disordered systems, on the other hand, $d\sigma/dT<0$ is usually not observed.  However, careful studies of $\sigma(n_s,T)$ ($n_s$ -- the electron density) and glassy charge dynamics in a 2DES in Si have provided ample evidence for the MIT and for the importance of  long-range Coulomb interactions also in these systems \cite{DP-CIQPT}.  The following key questions thus arise: (1) What is the nature of the MIT in a high-disorder 2DES with interactions?  More precisely, is it dominated by disorder, or is it the same as the MIT in a low-disorder 2DES, which is believed to be driven by interactions?  (2) What is the effect of the range of electron-electron interactions on the MIT in a high-disorder 2DES?  

Here we report a study of $\sigma(n_s,T)$ in high-disorder 2DES in Si MOSFETs, which demonstrates scaling behavior consistent with the existence of a quantum phase transition (QPT).  Measurements were done on devices in which the long-range part of the Coulomb interaction is screened by the metallic gate.  Scaling analysis was also performed on another sample of the same type, studied previously \cite{Snezana2002,Jan2006,Jan2007-1,Jan2007-2,Jaroszynski2009466,Lin2012}, but in which the electron-electron interaction is long range.  The comparison of our results to those on low-disorder systems provides clear evidence that sufficiently strong disorder changes the universality class of the MIT.  We also find that, in such a disorder-dominated transition, the range of the Coulomb interactions does not appear to affect the critical exponents.

The use of a nearby metallic gate or ground plane to limit the range of the Coulomb interactions between charge carriers in 2D systems is a well-known technique that has been explored both theoretically (see, \textit{e.g.}, \cite{Peeters1984,Widom1988,Hallam1996,Sushkov2009,Skinner-Shk,Skinner2010,Fregoso2013}) and experimentally, \textit{e.g.} in the investigation of the melting of the Wigner crystal formed by electrons on a liquid He surface \cite{Mistura1997}.  In the context of the
2D MIT, it has been used to explore the role of Coulomb interactions in the metallic \cite{Ho2008} and insulatorlike \cite{Huang2014} regimes of a 2D hole system (2DHS) in ``clean'', \textit{i.e.} low-disorder AlGaAs/GaAs heterostructures  and in the metallic regime of low-disorder Si MOSFETs \cite{Tracy}.  In contrast, we report on the screening by the metallic gate in a high-disorder 2D system.  Our conclusions are based on $\sigma(n_s, T)$ behavior on \emph{both} metallic and insulating sides of the MIT.

The metallic gate at a distance $d$ from the 2DES creates an image charge for each electron, modifying the Coulomb interaction from $\sim1/r$ to $\sim[1/r-1/\sqrt{r^2+4d^2}]$.  When the mean carrier separation $a=(\pi n_s)^{1/2}\gg d$, this potential falls off in a dipolelike fashion, as $\sim 1/r^3$.  Therefore, in Si MOSFETs, the range of the electron-electron Coulomb interactions can be changed by varying the thickness of the oxide $d_{ox}=d$.  Our study was performed on two sets of Si MOSFETs that were fabricated simultaneously using the 0.25-$\mu$m Si technology \cite{Taur},
the only difference being the value of $d_{ox}$.  In ``thick-oxide'' samples,  $d_{ox}=50$~nm, comparable to that in other Si MOSFETs used in the vast majority of studies of the 2D MIT \cite{2DMIT-review_2001, 2DMIT-review_2004, 2DMIT-review_2010,DP-CIQPT}.  In the low-$n_s$ regime of interest near the MIT, the corresponding $5.3\lesssim d/a \leq 8.0$.  On the other hand, in our ``thin-oxide'' devices with $d_{ox}=6.9$~nm, substantial screening by the gate is expected in the scaling regime of $n_s$ near the MIT, where $0.7\lesssim d/a \lesssim1.0$.  For comparison, in other ground-plane screening studies, $0.8\leq d/a \leq 1.8$ in Ref.~\cite{Tracy}, $1.1\lesssim d/a\leq 5$ in Ref.~\cite{Huang2014}, and $2\leq d/a\leq 19$ in Ref.~\cite{Ho2008}.

The samples were rectangular n-channel (100)-Si MOSFETs with poly-Si gates, self-aligned ion-implanted contacts, and oxide charge $N_{ox} \approx (1-1.5)\times 10^{11}$~cm$^{-2}$.  We focus on two samples that are representative of the two sets of devices: sample B$_{\mathrm{thin}}$, with $d_{ox}=6.9$~nm, substrate doping $N_a \sim 5 \times 10^{17}$~cm$^{-3}$, and dimensions $L \times W=2 \times 50$~$\mu$m$^2$ ($L$ -- length, $W$ -- width); sample A1 with $d_{ox}=50$~nm, $N_a \sim 2 \times 10^{17}$~cm$^{-3}$, and $L \times W=1 \times 90$~$\mu$m$^2$ \cite{Snezana2002,Lin2012}.  In analogy with previous studies on thick-oxide devices \cite{Snezana2002,Jan2006,Jan2007-1,Jan2007-2,Jaroszynski2009466,Lin2012}, the substrate (back-gate) bias of $-2$~V was applied, resulting in a 4.2~K peak mobility $\mu_{peak}$ of $\sim 0.04$~m$^2$/Vs and $\sim 0.06$~m$^2$/Vs for B$_{\mathrm{thin}}$ and A1, respectively.  Such low values of $\mu_{peak}$ reflect the presence of a large amount of disorder.  Detailed measurements were performed on sample B$_{\mathrm{thin}}$; the previously obtained data on A1 \cite{Snezana2002} were also analyzed.

$\sigma$ was measured using a standard two-probe ac method
at $\sim 11$~Hz with an ITHACO 1211 current preamplifier and a SR7265 lock-in amplifier in a $^3$He system (base $T=0.24$~K).  The contact resistances and the contact noise were determined to be negligible relative to those of the 2DES, as described in Ref.~\cite{Snezana2002}.  The excitation voltage $V_{exc}$ was
constant and low enough ($5-10~\mu$V) to ensure that the conduction was Ohmic. A precision dc voltage standard (EDC MV116J) was used to apply the gate voltage $V_g$, which controls $n_s$: $n_s(10^{11}$cm$^{-2})=31.25(V_g[$V$]-1.48)$ for sample B$_{\mathrm{thin}}$.  Similar to studies of thick-oxide devices \cite{Snezana2002,Jan2006,Jan2007-1,Jan2007-2,Jaroszynski2009466,Lin2012}, $n_s$ was varied at $T\approx 20$~K \cite{EF-note}, followed by cooling  
to a desired $T$ with a fixed $n_s$.  $\sigma$ was measured as a function of time, up to several hours at the lowest $n_s$ and $T$.  Some $V_g$ sweeps using a HP3325B function generator  were also performed to verify that $T\approx 20$~K was (a) high enough for the 2DES to be in a thermal equilibrium, as there were no visible relaxations, and (b) low enough for the background potential (disorder) to remain unchanged, as evidenced by the reproducible fluctuations of $\sigma(V_g)$ at low $T$.  The study of fluctuations with $V_g$ or with time, however, is beyond the scope of this work.  Here we focus instead on the behavior of the average conductivity $\langle\sigma\rangle$ \cite{Mirlin}. 

\begin{figure}
\includegraphics[width=8.2cm]{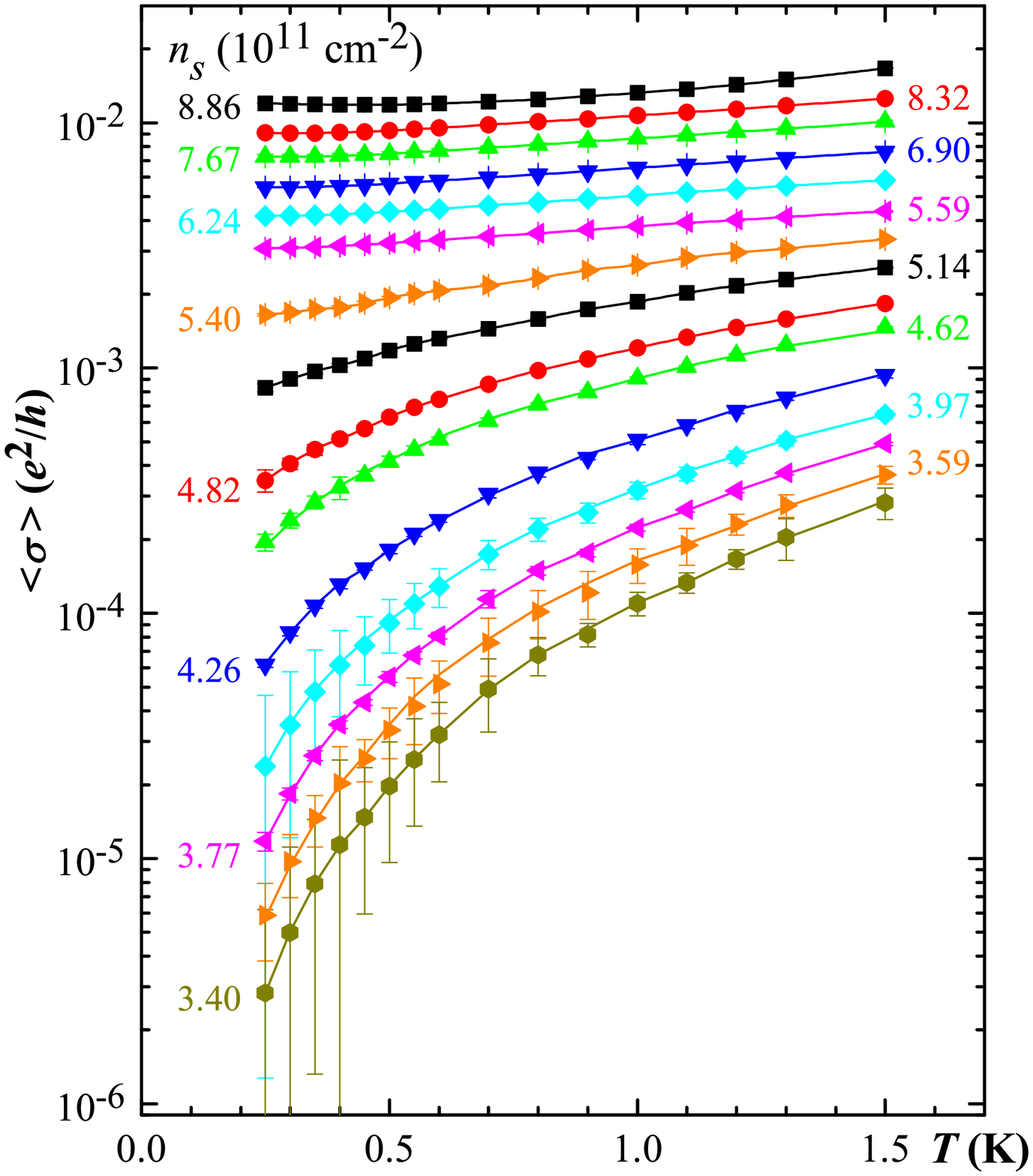}\llap{\parbox[b]{1.0in}{\textbf{(a)}\\\rule{0ex}{0.5in}}}
\includegraphics[width=8.2cm]{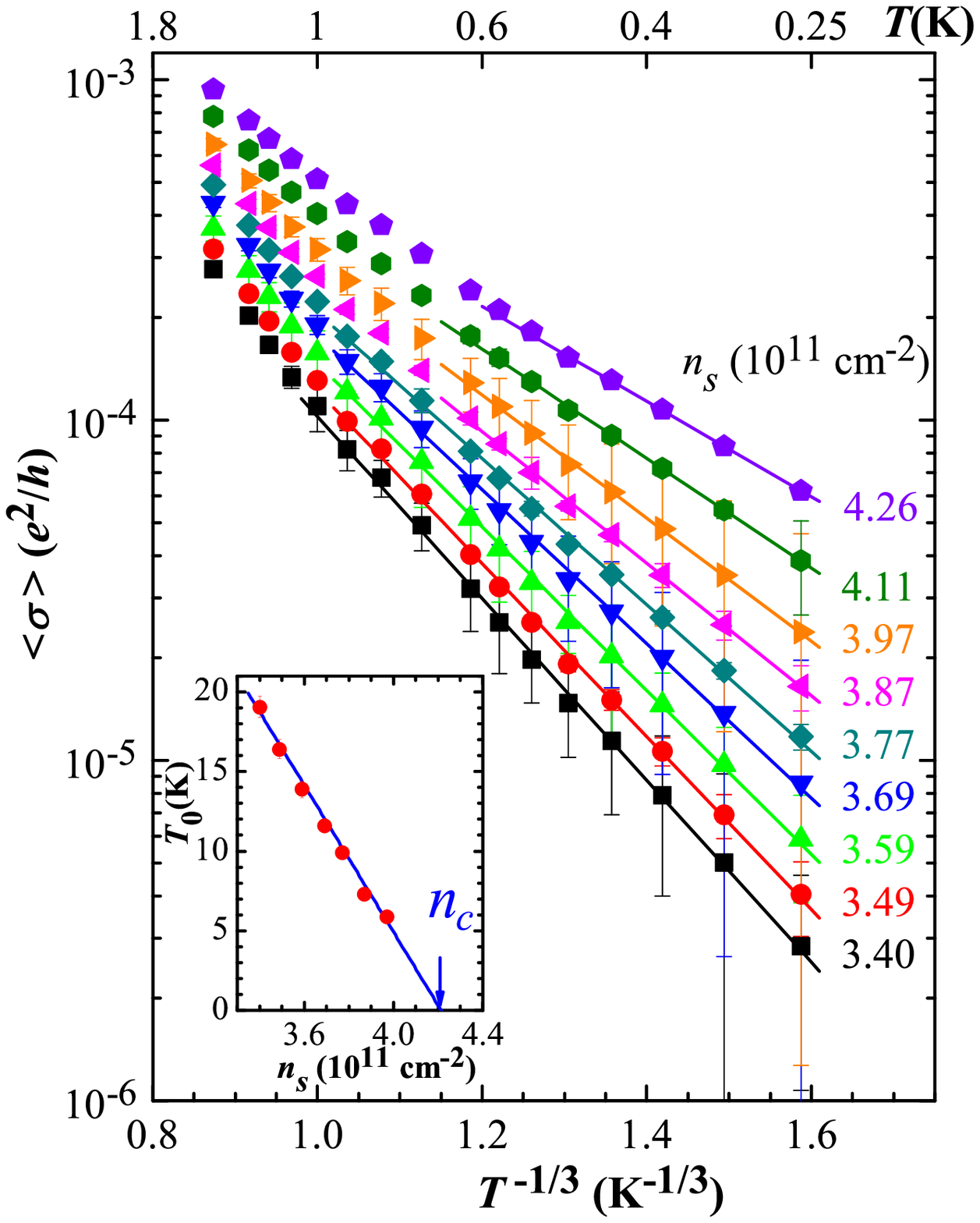}\llap{\parbox[b]{1.0in}{\textbf{(b)}\\\rule{0ex}{3.5in}}}
\caption {(Color online)  Sample B$_{\mathrm{thin}}$.  (a) Conductivity $\langle\sigma\rangle$ \textit{vs.} $T$ for different $n_s$, as shown.  $n_s$ was varied at high $T\approx 20$~K. 
Solid lines guide the eye.  All data are in the regime of $T\ll T_{F}$ ($T_{F}$ -- Fermi temperature \cite{EF-note}) and $k_{F}l \ll 1$ ($k_{F}$ -- Fermi wave vector, $l$ -- mean free path). 
(b) $\langle\sigma\rangle$ \textit{vs.} $T^{-1/3}$ for several $n_s$ in the insulating regime.  
The solid lines are fits to $\langle\sigma\rangle\propto\exp[-(T_0/T)^{1/3}]$.
Inset: $T_0$ \textit{vs.} $n_s$ with a linear fit, and an arrow showing $n_c$.  Only $n_s$ with the activation energies $E_A(T)=T_{0}^{1/3}T^{2/3} \gtrsim 0.6$~K were used in the fit.
In both (a) and (b), the error bars show the size of the fluctuations with time.}\label{fig:RvsT}
\end{figure}

Figure~\ref{fig:RvsT}(a) shows $\langle\sigma\rangle$ as a function of $T$ for different $n_s$ near the MIT, as discussed below.
In general, the behavior of $\langle\sigma(n_s, T)\rangle$ is similar to that in thick-oxide devices \cite{Snezana2002,Lin2012}, although the absolute values of $\langle\sigma\rangle$ for the same $n_s$ and $T$ are lower here.  We note that the mere decrease of $\sigma$ with decreasing $T$ (\textit{i.e.} $d\langle\sigma\rangle/dT >0$) at a given $n_s$ does not necessarily imply the existence of an insulating state ($\langle\sigma(T=0)\rangle=0$).  Indeed, the existence of a 2D metal with $d\langle\sigma\rangle/dT >0$ has been already demonstrated in three different types of 2DES in Si MOSFETs: 1) in the presence of scattering by disorder-induced local magnetic moments, both in zero magnetic field ($B=0$)\cite{Feng2001} and in parallel $B$ \cite{Eng2002}; 2) in low-disorder samples in parallel $B$ \cite{Jan2004}; and 3) in high-disorder, thick-oxide samples ($B=0$) \cite{Snezana2002}.  Therefore, $n_c$, the critical density for the MIT, in our high-disorder, thin-oxide system [Fig.~\ref{fig:RvsT}(a)] also has to be determined from the fits to $\langle\sigma(n_s, T)\rangle$ on both metallic and insulating sides of the transition.  
%

For the lowest $n_s$ and $T$, the best fit to the data is obtained with $\langle\sigma\rangle\propto\exp[-(T_0/T)^{1/3}]$ [Fig.~\ref{fig:RvsT}(b)], 
which corresponds to the 2D Mott variable-range hopping (VRH).  The vanishing of the activation energy, as extrapolated from the insulating phase, is often used as a criterion to determine $n_c$ (see, \textit{e.g.}, Refs. \cite{Pudalov-nc,Shashkin-nc,Snezana2002,Jan2002,Jan2004}).
Here the extrapolation of $T_0(n_s)$ to zero (Fig.~\ref{fig:RvsT}(b) inset) yields $n_c = (4.2\pm0.2)\times 10^{11}$~cm$^{-2}$. 

\begin{figure}
\includegraphics[width=8.1cm]{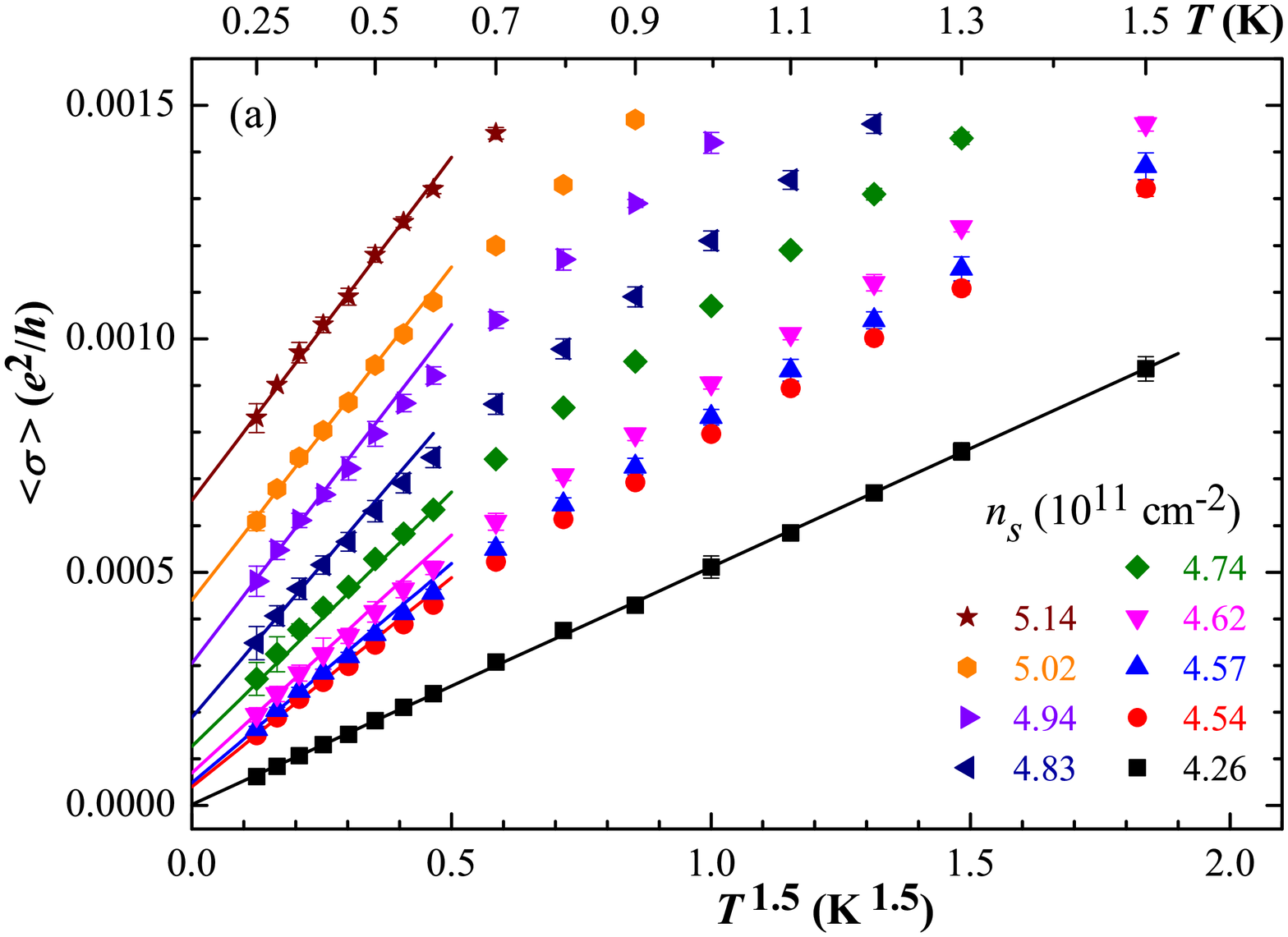}
\includegraphics[width=8.1cm]{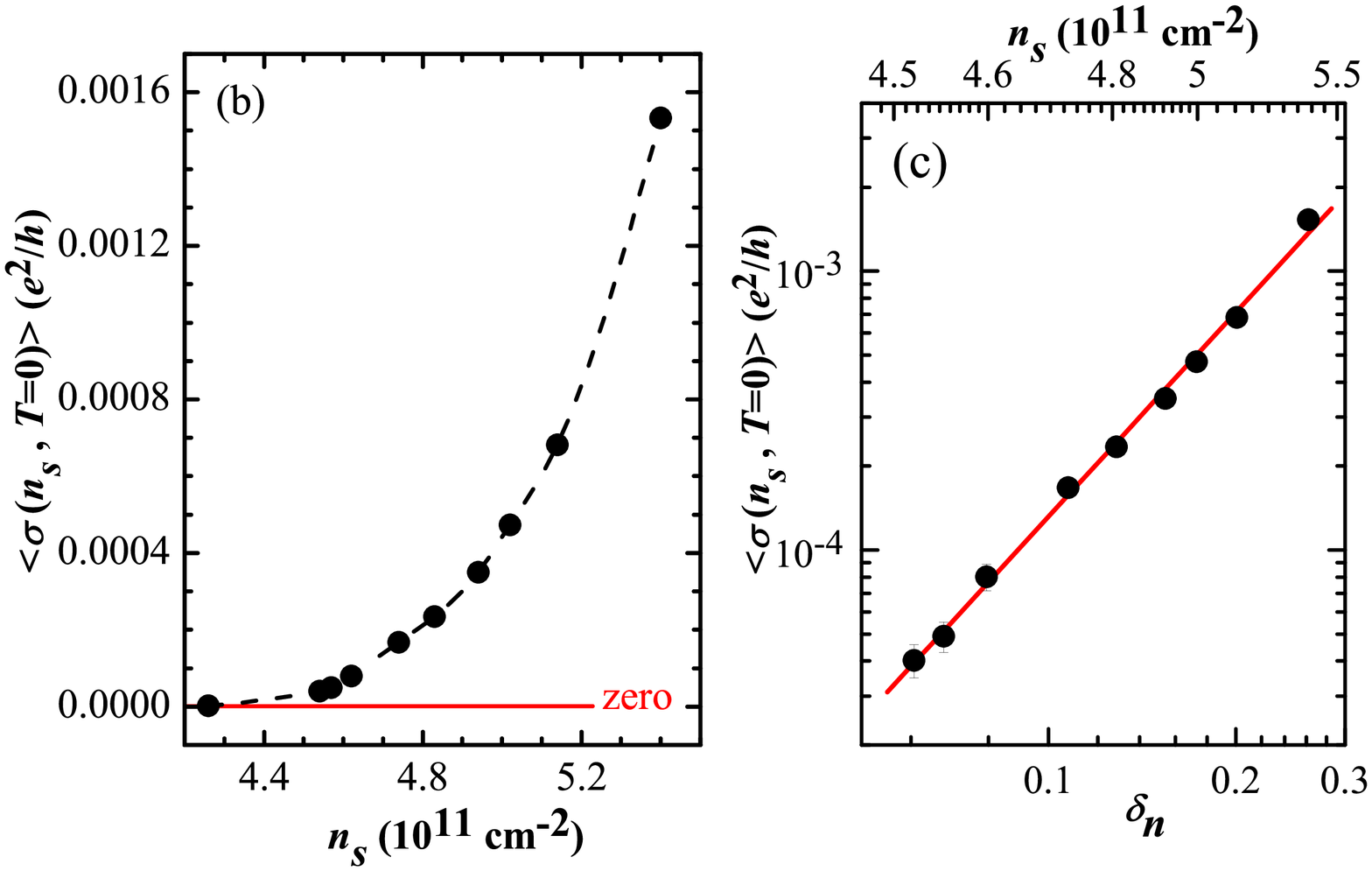}
\caption {(Color online) Sample B$_{\mathrm{thin}}$.  (a) $\langle\sigma\rangle$ \textit{vs.} $T^{1.5}$ for a few $n_s\geqslant n_c$, as shown.
The solid lines are linear fits.  For $n_s= 4.26\times10^{11}$~cm$^2$, $\langle\sigma(T=0)\rangle=0$, \textit{i.e.} $\langle\sigma(n_c,T)\rangle\propto T^x$ with $x=1.5\pm 0.1$.  (b) $\langle\sigma(n_s, T=0)\rangle$ \textit{vs.} $n_s$.  The dashed line guides the eye.  (c) $\langle\sigma(n_s, T=0)\rangle$ \textit{vs.} $\delta_n=(n_s-n_c)/n_c$, the distance from the MIT.  The solid line is a fit with the slope equal to the critical exponent $\mu=2.7\pm 0.3$. }\label{fig:metallic}
\end{figure}

For $n_s>n_c$, the low-$T$ data are best described by the metallic ($\langle\sigma(T=0)\rangle>0$) power law $\langle\sigma(n_s,T)\rangle=\langle\sigma(n _s, T=0)\rangle + b(n_s)T^{1.5}$ [Fig.~\ref{fig:metallic}(a)].  The same $T^{3/2}$ non-Fermi-liquid correction was observed in the metallic, glassy phase ($n_c<n_s<n_g$; $n_g$ - glass transition density) of both thick-oxide, high-disorder samples at $B=0$ \cite{Snezana2002} and low-disorder 2DES in parallel $B$ \cite{Jan2004}, consistent with theoretical predictions \cite{Dalidovich,Sachdev-2012}.  This simple and precise form of $\langle\sigma(T)\rangle$ allows a reliable extrapolation to $T=0$ [Fig.~\ref{fig:metallic}(a)].  The extrapolated $\langle\sigma(T=0)\rangle$
go to zero at $n_s\approx 4.26\times10^{11}$~cm$^2$ [Fig.~\ref{fig:metallic}(b)], in agreement with the $n_c$ value obtained from the VRH fit.  Moreover, a simple power-law $T$ dependence $\langle\sigma(n_c,T)\rangle\propto T^x$ found here (Fig.~\ref{fig:metallic}(a); $x=1.5$), is consistent with the one expected in the quantum critical region of the MIT based on general arguments \cite{Belitz-RMP}.  Likewise, the power-law behavior $\langle\sigma(n_s, T=0)\rangle\propto\delta_{n}^{\mu}$ [Fig.~\ref{fig:metallic}(c)] is in agreement with theoretical expectations near a QPT, such as the MIT.  The critical exponent $\mu=2.7\pm 0.3$.

\begin{figure}
\includegraphics[width=8.1cm]{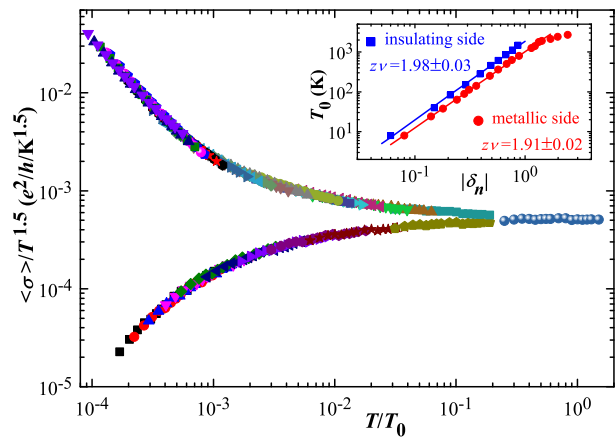}
\caption {(Color online) Scaling of $\langle\sigma\rangle/\langle\sigma_{c}\rangle\propto\langle\sigma\rangle/T^{x}$, $x=1.5$, with $T$ 
for sample B$_{\mathrm{thin}}$ ($d_{ox}=6.9$~nm).  Different symbols correspond to $n_s$ from $3.40\times10^{11}$~cm$^{-2}$ to $6.70\times10^{11}$~cm$^{-2}$; $n_c= 4.26\times10^{11}$~cm$^2$.  It was possible to scale the data below about 1.5~K.  Inset: $T_0$ \textit{vs.} $\delta_n$.  The lines are fits with slopes $z\nu=1.98\pm0.03$ and $z\nu=1.91\pm0.02$ on the insulating and metallic sides, respectively.}\label{fig:scalingthin}
\end{figure}

In addition, very general considerations have suggested \cite{Belitz-RMP} that the conductivity near the MIT can be described by a scaling form $\langle\sigma(n_s, T)\rangle=\langle\sigma_c(T)\rangle f(T/\delta_{n}^{z\nu})$, where $z$ and $\nu$ are the dynamical and correlation length exponents, respectively, and the critical conductivity $\langle\sigma_c\rangle=\langle\sigma(n_s=n_c, T)\rangle\propto T^x$. 
Figure~\ref{fig:scalingthin} shows that, in the vicinity of $n_c$, all $\langle\sigma(n_s, T)\rangle/\langle\sigma_{c}(T)\rangle\propto\langle\sigma(n_s, T)\rangle/T^{1.5}$ collapse onto the same function $f(T/T_0)$ with two branches: the upper one for the metallic side of the transition and the lower one for the insulating side.  As expected for a QPT, the scaling parameter $T_{0}$ is the same, power-law function of $\delta_n$ on both sides of the transition, $T_0\propto|\delta_n|^{z\nu}$  (Fig.~\ref{fig:scalingthin} inset), with $z\nu\approx2.0$ within experimental error.

%
%
From standard scaling arguments \cite{Belitz-RMP}, it follows that the critical exponent $\mu$ can be determined not only from extrapolations of $\langle\sigma(n _s, T)\rangle$ to $T=0$ [Fig.~\ref{fig:metallic}(c)], but also from $\mu=x(z\nu)$ based on all data taken at all $T$ and values of $n_s$ for which scaling holds.  Indeed, using $x=1.5\pm 0.1$ (Figs.~\ref{fig:metallic}(a) and \ref{fig:scalingthin}) and $z\nu\approx 2$ (Fig.~\ref{fig:scalingthin} inset),
we find the value $\mu=x(z\nu)= 3.0\pm0.3$ that is in excellent agreement with $\mu=2.7\pm 0.3$ found from the $T=0$ extrapolation of $\langle\sigma(n _s, T)\rangle$.  This confirms the consistency of the analysis.

\begin{figure}
\includegraphics[width=8.1cm]{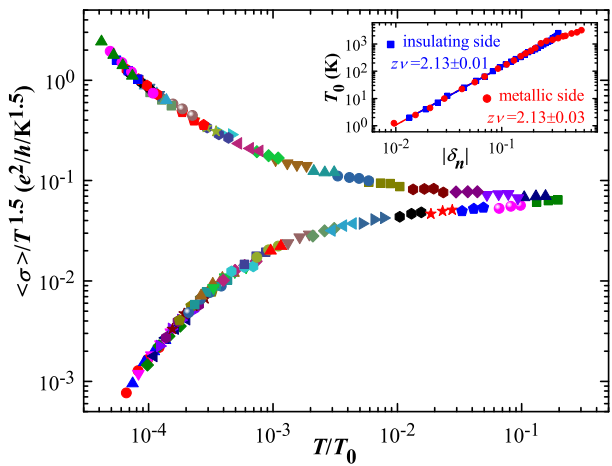}
\caption {(Color online) Scaling of $\langle\sigma\rangle/\langle\sigma_{c}\rangle\propto\langle\sigma\rangle/T^{x}$, $x=1.5$, with $T$ for sample A1 ($d_{ox}=50$~nm).  Different symbols correspond to $n_s$ from $3.45\times10^{11}$~cm$^{-2}$ to $8.17\times10^{11}$~cm$^{-2}$; $n_c= 5.22\times10^{11}$~cm$^2$.  It was possible to scale the data below $\sim 0.3$~K down to the lowest $T=0.13$~K.  Inset: $T_0$ \textit{vs.} $\delta_n$.  The lines are fits with slopes $z\nu=2.13\pm0.01$ and $z\nu=2.13\pm0.03$ on the insulating and metallic sides, respectively.}\label{fig:scalingthick}
\end{figure}

In a similar way, we analyze $\langle\sigma(n_s, T)\rangle$ near $n_c= 5.22\times10^{11}$~cm$^2$ in a thick-oxide, high-disorder MOSFET \cite{Snezana2002}, in which the electron-electron interaction is long range.  Figure~\ref{fig:scalingthick} demonstrates that, near $n_c$, the $\langle\sigma(n_s, T)\rangle/\langle\sigma_{c}(T)\rangle\propto\langle\sigma(n_s, T)\rangle/T^{1.5}$ data exhibit dynamical scaling, a signature of the QPT, also in this system.  The scaling parameter $T_0\propto|\delta_n|^{z\nu}$  (Fig.~\ref{fig:scalingthick} inset), with $z\nu\approx2.1$ within experimental error.  Therefore, the critical exponents are the same as those in thin-oxide, high-disorder samples, and thus not sensitive to the range of the Coulomb interactions.  
%

The critical exponents have been summarized in Table~\ref{table1}.  The Table also shows critical exponents obtained in 2DESs with much lower disorder (\textit{i.e.} high $\mu_{peak}$) \cite{Kravchenko1995, Sarachik1996, Popovic1997, Pudalov1998, Feng1999, Jan2004, Ted-Belitz-2013, Fletcher2001}, including those in which scattering by local magnetic moments dominates \cite{Feng2001,Eng2002}.  The values obtained in low parallel $B$ (\textit{i.e.} $B$ not high enough to fully spin polarize the 2DES \cite{spin-polarization1, spin-polarization2}) are also included, where available.  It is apparent that such low fields do not seem to affect any of the critical exponents.  On the other hand, we find a major difference between $z\nu\approx 2.0$ in our low-$\mu_{peak}$ devices and, consistently lower, $z\nu=1.0-1.7$ in high-$\mu_{peak}$ 2DES \cite{Kravchenko1995,Sarachik1996,Popovic1997, Pudalov1998, Feng1999, Feng2001, Ted-Belitz-2013}.  This result indicates that sufficiently strong disorder changes the nature of the MIT from interaction-driven in high-$\mu_{peak}$ samples to disorder-dominated in low-$\mu_{peak}$ 2DES.  In such a disorder-dominated MIT, it is plausible that the range of the Coulomb interactions does not seem to play a major role.  The possibility of a disorder-dominated 2D MIT has been demonstrated theoretically \cite{Punnoose} for both long-range and short-range interactions.  Although there is currently no microscopic theory that describes the detailed properties of the observed MIT, it is interesting that in the available theories \cite{Belitz-RMP, Punnoose}, the range of the Coulomb interactions does not play a significant role.  We also note that percolation models \cite{percolation} cannot describe our findings, \textit{e.g.} the 2D percolation $\mu\simeq 1.3$, as opposed to the much larger experimental $\mu\simeq 3$ (Table~\ref{table1}).  Interestingly,  the same large $\mu\simeq 3$ was observed in a high-$\mu_{peak}$ 2DES ($z\nu\approx 1.3$) in the presence of scattering by local magnetic moments \cite{Feng2001}.  Therefore, unlike $z\nu$, the exponent $x$ seems to be more sensitive to the type (\textit{e.g.} magnetic \textit{vs.} nonmagnetic), rather than to the amount of disorder.

\begin{table}
\Large
\renewcommand\arraystretch{2.0}
\centering
\caption{Critical exponents $x$, $z\nu$, $\mu$ (determined from $\langle\sigma(n_s, T=0)\rangle\propto\delta_{n}^{\mu}$), and $\mu=x(z\nu)$ for 2D electron systems in Si MOSFETs with different disorder.  The 4.2~K peak mobility $\mu_{peak}$[m$^2$/Vs] is a rough measure of the amount of disorder. $d_{ox}$[nm] is the oxide thickness, $n_c[10^{11}$~cm$^{-2}$] is the critical carrier density for the MIT in zero magnetic field.  In low parallel $B$, $[n_c(B)/n_c(0)-1]\propto B^{\beta}$  with $\beta=1.0\pm 0.1$ for both low-disorder samples \cite{Dolgopolov1992,Sakr,Shashkin-nc,Jan2004} and those in which scattering by local magnetic moments dominates \cite{Eng2002}.  ``--'' indicates that the data are either insufficient or unavailable.  
\vspace{6pt}}   
\newsavebox{\tableboxtwo}
\begin{lrbox}{\tableboxtwo}
\begin{threeparttable}
\begin{tabular}{|c|c|c|c|c|c|c|}
\hline 
 &\multicolumn{2}{c|}{High-disorder system}&\multicolumn{2}{c|}{\multirowcell{2}{Special disorder:\\ local magnetic moments}}&\multicolumn{2}{c|}{\multirowcell{2}{Low-disorder\\ system}}\\ 
\cline{2-3}
 &\makecell{thin oxide}&\makecell{thick oxide}& \multicolumn{2}{c|}{} & \multicolumn{2}{c|}{}\\
    \hline
    $\mu_{peak}$  & 0.04 & 0.06 &\multicolumn{2}{c|}{$\sim 1$} & \multicolumn{2}{c|}{$\sim 1-3$}\\
    \hline
    $d_{ox}$ & 6.9 & 50 &\multicolumn{2}{c|}{43.5} &\multicolumn{2}{c|}{40-600} \\
    \hline
     & $B=0$ &$B=0$  & $B=0$ \cite{Feng2001} & $B\neq 0$ \cite{Eng2002} & $B=0$ & $B\neq 0$ \\
      \hline
    $n_c$  & $4.2\pm0.2$ & $5.0\pm0.3$ & 0.5-1 & $[\frac{n_c(B)}{n_c(0)}-1]\propto B$ & $\sim 1$ & $[\frac{n_c(B)}{n_c(0)}-1]\propto B$ \\
     \hline
    $x$ & $1.5\pm0.1$ & $1.5\pm0.1$ & $2.6\pm 0.4$ & $2.7\pm 0.4$  & -- & $1.5\pm 0.1$ \cite{Jan2004}\\
    \hline
    $z\nu$ & $2.0\pm0.1$ & $2.1\pm0.1$ & $1.3\pm 0.1$ & $0.9\pm 0.3$ & $1.0-1.7$ \cite{Kravchenko1995,Sarachik1996,Popovic1997, Pudalov1998, Feng1999,Ted-Belitz-2013} & --\\
    \hline
    $\mu$ & $2.7\pm0.3$ & -- & $3.0\pm 0.1$ & $3.0\pm 0.1$ & 1-1.5 \cite{Fletcher2001} & $1.5\pm 0.1$ \cite{Jan2004}\\
    \hline
    $\mu=x(z\nu)$ & $3.0\pm0.3$ & $3.3\pm0.4$ & $3.4\pm 0.4$ & $2.4\pm 1$ & -- & --\\
\hline
\end{tabular}\label{table1}
\end{threeparttable}
\end{lrbox}
\resizebox{\columnwidth}{!}{\usebox{\tableboxtwo}} 
\end{table}

Our study demonstrates the critical behavior of $\sigma$ consistent with the existence of a metal-insulator quantum phase transition in a highly disordered 2DES in Si MOSFETs.  The results strongly suggest that, in contrast to the MIT in a low-disorder 2DES, the MIT reported here is dominated by disorder.  We have also established that the range of the Coulomb interactions does not seem to affect the properties, \textit{i.e.} the critical exponents, of such a disorder-dominated MIT.  On the other hand, the effect of the range of electron-electron interactions on the critical behavior of a low-disorder 2DES remains an open question.

The authors acknowledge the IBM T. J. Watson Research Center for fabricating the devices and V. Dobrosavljevi\'{c} for useful discussions.  This work was supported by NSF grants DMR-0905843, DMR-1307075, and the National High Magnetic Field Laboratory through NSF Cooperative Agreement DMR-1157490 and the State of Florida.


\end{document}